\pdfoutput=1
\documentclass[sigconf,authorversion=true]{acmart}

\AtBeginDocument{%
  \providecommand\BibTeX{{%
    \normalfont B\kern-0.5em{\scshape i\kern-0.25em b}\kern-0.8em\TeX}}}

\acmConference[KDD '24]{ACM Conference}{August 25--29, 2024}{Barcelona, Spain}%
\acmBooktitle{Proceedings of the 30th ACM SIGKDD Conference on Knowledge Discovery and Data Mining (KDD '24), August 25--29, 2024, Barcelona, Spain}
\acmDOI{10.1145/3637528.3671896}
\acmISBN{979-8-4007-0490-1/24/08}

\usepackage[utf8]{inputenc} 
\usepackage[T1]{fontenc}    
\usepackage{url}            
\usepackage{nicefrac}       
\usepackage{microtype}      
\usepackage{xcolor}         
\usepackage{bm}
\usepackage{bbm}
\usepackage{todonotes}
\usepackage{algorithm}
\usepackage{algpseudocode}
\usepackage{multicol}
\usepackage{caption}
\usepackage{multirow}
\usepackage{tabularx}
\usepackage{makecell}
\usepackage{booktabs}           
\usepackage{amsfonts}           
\usepackage{graphicx}           
\usepackage{duckuments}         
\usepackage{bold-extra}
\usepackage{balance}

\usepackage{caption}
\usepackage{subcaption}
\usepackage{svg}

\newcommand\numberthis{\addtocounter{equation}{1}\tag{\theequation}}

\DeclareMathOperator*{\argmax}{arg\,max}

\newcommand{\R}{\mathbb{R}}

\newcommand{\G}{\mathcal{G}}

\newcommand{\V}{\mathcal{V}}

\newcommand{\edges}{\mathcal{E}}

\newcommand{\graph}{\G=(\V,\edges)}

\newcommand{\params}{\bm{\theta}}

\newcommand{\loss}{\ell}
\newcommand{\Loss}{\mathcal{L}}

\begin{document}

\title[Evading Community Detection via Counterfactual Neighborhood Search]{Evading Community Detection via\\ Counterfactual Neighborhood Search}

\author{Andrea Bernini}
\orcid{0009-0009-2870-2673}
\affiliation{%
  \institution{Sapienza University of Rome}
  \country{Italy}
}
\email{andreabe99@gmail.com}

\author{Fabrizio Silvestri}
\orcid{0000-0001-7669-9055}
\affiliation{%
  \institution{Sapienza University of Rome}
   \country{Italy}
   }
\email{fsilvestri@diag.uniroma1.it}

\author{Gabriele Tolomei}
\orcid{0000-0001-7471-6659}
\affiliation{%
  \institution{Sapienza University of Rome}
   \country{Italy}
   }
\email{tolomei@di.uniroma1.it}

\renewcommand{\shortauthors}{Andrea Bernini, Fabrizio Silvestri, \& Gabriele Tolomei}

\begin{abstract}
Community detection techniques are useful for social media platforms to discover tightly connected groups of users who share common interests.
However, this functionality often comes at the expense of potentially exposing individuals to privacy breaches by inadvertently revealing their tastes or preferences. 
Therefore, some users may wish to preserve their anonymity and opt out of community detection for various reasons, such as affiliation with political or religious organizations, without leaving the platform.\\
In this study, we address the challenge of \textit{community membership hiding}, which involves strategically altering the structural properties of a network graph to prevent one or more nodes from being identified by a given community detection algorithm. 
We tackle this problem by formulating it as a constrained counterfactual graph objective, and we solve it via deep reinforcement learning. 
Extensive experiments demonstrate that our method outperforms existing baselines, striking the best balance between accuracy and cost.
\end{abstract}





\maketitle

\section{Introduction}
\label{sec:intro}

Identifying \textit{communities} is crucial for understanding the intricacies of complex graph structures like social networks \cite{fortunato2010pr}.
This is typically achieved by \textit{community detection} algorithms, which group nodes based on shared characteristics or interactions, shedding light on the underlying organization and dynamics of the network.

The successful detection of communities in complex network graphs is useful in several application domains \cite{karastas2018ibigdelft}.
For instance, the insights gained from accurately identifying communities can significantly impact business strategies, leading to better monetization opportunities through targeted advertising \cite{mosadegh2011ajbmr}.

However, community detection algorithms also raise concerns on individual privacy and data protection. 
In some cases, certain nodes within the network might prefer not to be identified as part of a particular community. These nodes could be associated with sensitive or private groups (e.g., political or religious organizations) or may wish to protect their anonymity for personal reasons.
An option for these users would be to leave the platform, but such a decision might be too extreme.
A more flexible approach would allow users to opt out of community detection while staying on the platform. This strategy strikes the optimal balance between preserving privacy and maximizing the utility of community detection.

Motivated by the need above, in this paper, we address the intriguing challenge of \textit{community membership hiding}.


Drawing inspiration from \textit{counterfactual reasoning} \cite{tolomei2017kdd, tolomei2021tkde}, particularly in the realm of graph data \cite{lucic2022aistats}, we aim to provide users with personalized recommendations that safeguard their anonymity from community detection. To illustrate this with an example, consider a scenario in which a social network is equipped with a tool that offers guidance to its users on how to modify their connections to prevent them from being recognized as members of a particular community. For instance, a recommendation could take the form: "\textit{If you unfollow users X and Y, you will no longer be recognized as a member of community Z.}"

The core challenge of this problem lies in determining how to strategically modify the structural properties of a network graph, effectively excluding one or more nodes from being identified by a given community detection algorithm.
To the best of our knowledge, we are the first to formulate the community membership hiding problem as a constrained counterfactual graph objective. 
Furthermore, inspired by \cite{chen2022cikm}, we cast this problem into a Markov decision process (MDP) framework and we propose a deep reinforcement learning (DRL) approach to solve it. 

Our method works as follows. 
We start with a graph and a set of communities identified by a specific community detection algorithm whose inner logic can be unknown or undisclosed.
When given a target node within a community, we aim to find the optimal structural adjustment of the target node's neighborhood. This adjustment should enable the target node to remain concealed when the (same) community detection algorithm is reapplied to the modified graph.
We refer to this as \textit{community membership hiding task}, and we consider it successful when a predefined similarity threshold between the original community and the new community containing the target node is met. 
It is worth noting that while related, this task differs significantly from the \textit{community deception task} explored in existing literature \cite{deception_modularity_3}. 
Specifically, community deception aims to hide an entire community from community detection algorithms by rewiring connections of \textit{some} members within the target community. 
However, the community deception task, as defined by \citet{deception_modularity_3}, does not have a binary outcome, unlike our community membership hiding goal. 
Instead, the authors introduce a smooth measure, the \textit{Deception Score}, that combines three criteria for effective community masking: reachability, spreadness, and hiding.
While it might seem plausible to extend our method for community deception by running multiple membership hiding tasks for \textit{each} node in the target community, this straightforward strategy might be too aggressive due to our more stringent (i.e., binary) definition of deception goal. 
A more nuanced approach could involve leveraging the structural properties of each node in the community to mask (e.g., their degree) to cleverly select target nodes for membership hiding. Further exploration of this strategy is left for future research.

We validate our approach on five real-world datasets, and we demonstrate that it outperforms existing baselines using standard quality metrics. 
Notably, our method maintains its effectiveness even when used in conjunction with a community detection algorithm that was not seen during the training phase. 
We call this key property: \textit{transferability}.

Our main contributions are summarized below.
\begin{itemize}
    \item We formulate the community membership hiding problem as a constrained counterfactual graph objective. 
    \item We cast this problem within an MDP framework and solved it via DRL.
    \item We utilize a graph neural network (GNN) representation to capture the structural complexity of the input graph, which in turn is used by the DRL agent to make its decisions.
    \item We validate the performance of our method in comparison with existing baselines using standard quality metrics.
    \item We publicly release both the source code and the data utilized in this study to encourage reproducibility.\footnote{ \url{https://github.com/AndreaBe99/community_membership_hiding}}
\end{itemize}

The remainder of this paper is structured as follows. 
In Section~\ref{sec:related}, we review related work. Section~\ref{sec:background} contains background and preliminaries. 
We present our problem formulation in Section~\ref{sec:problem}. 
In Section~\ref{sec:method}, we describe our method, which we validate through extensive experiments in Section~\ref{sec:experiments}. 
Section~\ref{sec:discussion} discusses the implementation challenges along with the potential security and ethical impact of our method.
Finally, we conclude in Section~\ref{sec:conclusion}.

\section{Related Work}
\label{sec:related}

\noindent \textbf{\textit{Community Detection}.}
Community detection algorithms are essential tools in network analysis, aiming to uncover densely connected groups of nodes within a graph. They apply to various domains, such as social network analysis, biology, and economics.

These algorithms can be broadly categorized into two types: non-overlapping and overlapping community detection.

Non-overlapping community detection assigns each node to a single community, employing various techniques such as Modularity Optimization \cite{louvain_detection_alg}, Spectrum Analysis \cite{spectrum_analysis}, Random Walk \cite{walktrap_detection_alg}, or Label Propagation \cite{label_detection_alg}.
On the other hand, overlapping community detection seeks to represent better real-world networks where nodes can belong to multiple communities. Mature methods have been developed for this purpose, including NISE (Neighborhood-Inflated Seed Expansion) \cite{nise} and techniques based on minimizing the Hamiltonian of the Potts model \cite{Ronhovde_2009}. 
For a more comprehensive overview, we recommend consulting the work by \citet{community_detection_survey}.

\noindent \textbf{\textit{Community Deception}.}
As already discussed in Section~\ref{sec:intro} above, community deception is closely related to the community membership hiding problem we investigate in this work. 
Somehow, it is a specialization of community membership hiding, where the objective is to hide an entire community from community detection algorithms. 
In essence, this should involve performing multiple membership hiding tasks, one for each node within the community to be masked. 
However, this is a pretty raw simplification of the process since each node hiding task, for how we define it below, may, once solved, already achieve a satisfactory degree of community hiding (i.e., obtain a high Deception Score, according to \cite{deception_modularity_3}), indirectly concealing multiple nodes within the same community.

Community deception can serve various purposes, such as preserving the anonymity of sensitive groups of individuals in online monitoring scenarios, like social networks, or aiding public safety by identifying online criminal activities.
Yet, these techniques also pose risks, as malicious actors can use them to evade detection algorithms and operate covertly, potentially violating the law.

Several techniques exist for hiding communities, such as those based on the concept of Modularity. Notable examples in this category include the approach proposed by \citet{deception_modularity_1}, the DICE algorithm introduced by \citet{deception_modularity_2}, and the method developed by \citet{deception_modularity_3}. 
Other deception techniques are founded on the idea of Safeness, as defined by \citet{deception_modularity_3} and further explored by \citet{deception_safeness_2}, as well as the notion of Permanence used by \citet{deception_neural}.
An exhaustive summary of these methods is provided by \citet{deception_survey}.

\section{Background and Preliminaries}
\label{sec:background}
In this section, we briefly review the well-known community detection problem and utilize its definition as the basis for formulating the community membership hiding problem.

Let $\graph$ be an arbitrary (directed) graph, where 
$\V$ is a set of $n$ nodes ($|\V| = n$), and $\edges \subseteq \V \times \V$ is a set of $m$ edges ($|\edges| = m$).
Optionally, an additional set of $p$ node attributes  
may also be present. 
In such cases, each node $u\in \V$ is associated with a corresponding $p$-dimensional real-valued feature vector $\bm{x}_u \in \R^p$. 
Furthermore, the underlying link structure of $\G$ is represented using a binary adjacency matrix $\bm{A}\in \{0,1\}^{n\times n}$, where $\bm{A}_{u,v} = 1$ if and only if the edge $(u, v)\in \edges$, and it is $0$ otherwise.


The \textit{community detection} problem aims to identify clusters of nodes within a graph, called \textit{communities}.
Due to the intricate nature of the concept and its reliance on contextual factors, establishing a universally accepted definition for a "community" is challenging. 
Intuitively, communities exhibit strong intra-cluster connections and relatively weaker inter-cluster connections \cite{zhao2012jstor}.

More formally, in this work, we adhere to the definition widely used in the literature \cite{deception_modularity_3, deception_persistence_1, deception_rem_1, deception_neural}, and we consider a function $f(\cdot)$ that takes a graph as input and generates a partition of its nodes $\V$ into a set of non-empty, non-overlapping, communities $\{\mathcal{C}_1, \ldots, \mathcal{C}_k\}$ as output, i.e., $f(\G) = \{\mathcal{C}_1, \ldots, \mathcal{C}_k\}$, where $k$ is usually unknown. 
Within this framework, every node $u \in \V$ is assigned to \textit{exactly one} community. 
However, this can be extended to accommodate scenarios with overlapping communities, where each node can belong to multiple clusters.
Indeed, to represent node-community assignments, we can use a $k$-dimensional stochastic vector $\bm{c}_u$, where $c_{u,i} = P(u \in \mathcal{C}_i)$ measures the probability that node $u$ belongs to community $\mathcal{C}_i$, and $\sum_{i=1}^k c_{u,i} = 1$.
Eventually, we use the notation $i^{*}_u = \argmax_i(c_{u,i})$ to define the index of the community to which a specific node $u$ belongs based on the outcome of $f(\G)$. 
Note that in the case of hard node partitioning, the vector $\bm{c}_u$ has \textit{only one} non-zero entry, which equals $1$.
We leave the exploration of overlapping communities for future work.


Typically, community detection methods operate by maximizing a specific score that measures the intra-community cohesiveness (e.g., Modularity \cite{newman2006pnas}). 
However, this usually translates into solving NP-hard optimization problems. 
Hence, some convenient approximations have been proposed in the literature to realize $f(\cdot)$ in practice, e.g., Louvain \cite{louvain_detection_alg}, WalkTrap \cite{walktrap_detection_alg}, Greedy \cite{greedy_detection_alg}, InfoMap \cite{infomap_detection_alg}, Label Propagation \cite{label_detection_alg}, Leading~Eigenvectors \cite{eigenvectors_detection_alg}, Edge-Betweeness \cite{edge_detection_alg}, SpinGlass \cite{spinglass_detection_alg}.
Anyway, the rationale behind how communities are found is irrelevant to our task, and, hereinafter, we will treat the community detection technique $f(\cdot)$ as a "black box."

\section{Community Membership Hiding}
\label{sec:problem}

In a nutshell, community membership hiding seeks to allow a target node in a graph to avoid being identified as a member of a specific node cluster, as determined by a community detection algorithm. 
This objective is achieved by suggesting to the node in question how to strategically modify its connections with other nodes. 
Our primary focus is to change the graph's structure, represented by the adjacency matrix. While altering node features holds potential interest, that aspect is reserved for future work.  

\noindent{\textbf{\textit{Assumptions:}} Whoever runs our community membership hiding algorithm must be able to \textit{execute} the community detection algorithm $f(\cdot)$ even without access to its internal logic and possesses \textit{full knowledge} of the graph. Nevertheless, our method might work under more relaxed conditions (e.g., partial graph knowledge), as discussed in Section~\ref{subsec:implementation}.
We illustrate our approach in Figure~\ref{fig:node_deception_process}. 
\begin{figure}[htbp]
\includegraphics[width=0.75\columnwidth]{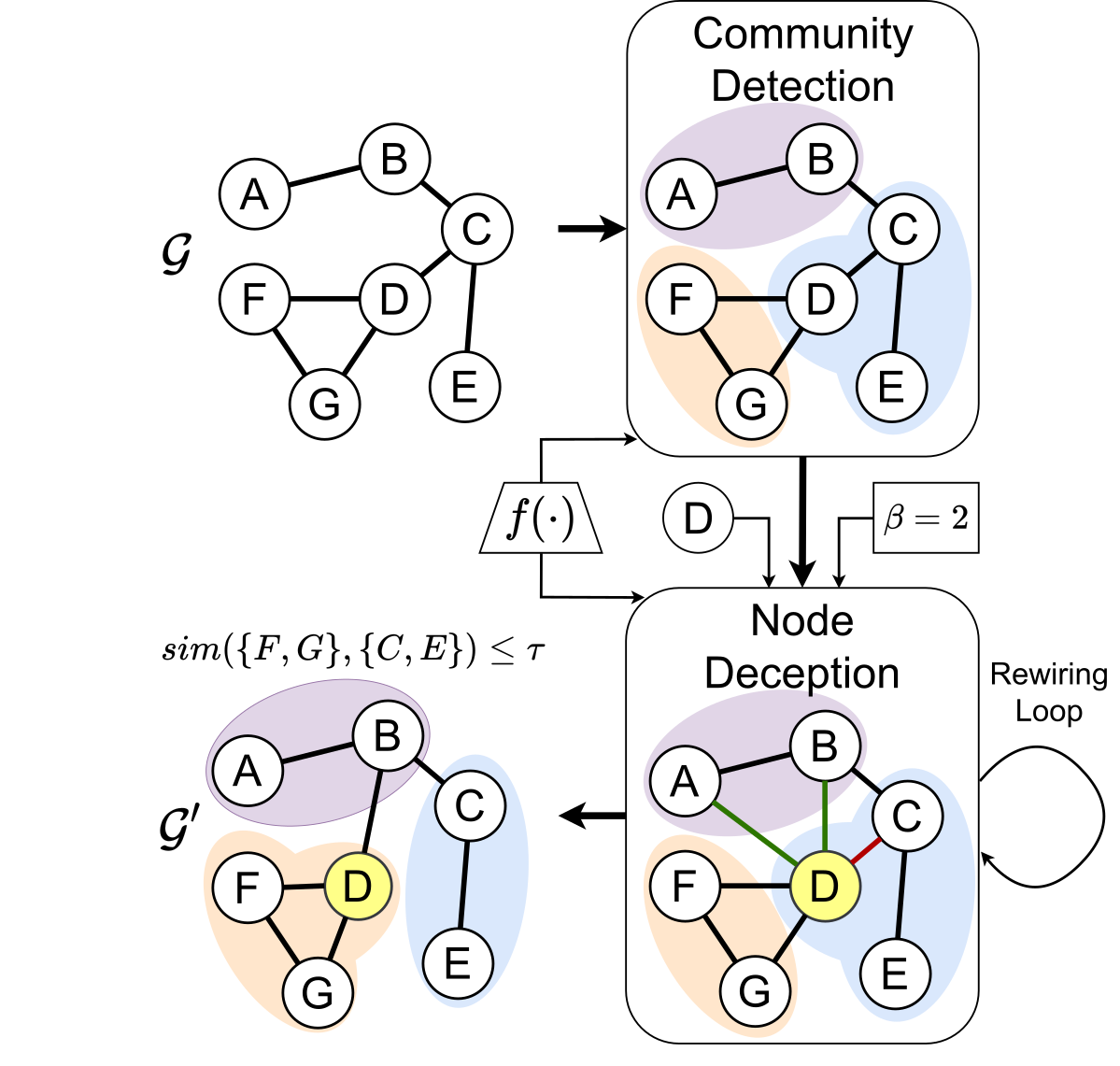}
\centering
\caption{Given a graph $\G$, a node $u$ (in this case $u=D$), a budget of actions $\beta$, and the set of communities identified by the community detection algorithm $f(\cdot)$ (including the community $\mathcal{C}_i$ to which the node belongs), \textit{community membership hiding} consists of adding inter-community edges $\edges_{u,i}^+$ (green edges), or removing intra-community edges $\edges_{u,i}^-$ (red edge), so that the value returned by the similarity function $sim(\cdot, \cdot)$, between the new community to which the node belongs after rewiring, and the original one, is lower than the $\tau$ constraint.}
\label{fig:node_deception_process}
\end{figure}

\subsection{Problem Formulation}
Let $\graph$ be a graph and $f(\mathcal{G}) = \{\mathcal{C}_1,\ldots,\mathcal{C}_k\}$ denote the community arrangement derived from applying a detection algorithm $f(\cdot)$ to $\mathcal{G}$.
Furthermore, suppose that $f$ has identified node $u\in \mathcal{V}$ as a member of the community $\mathcal{C}_i\in f(\G)$ -- i.e., $i^*_u = i$ -- denoted as $u\in \mathcal{C}_i$. 
The aim of community membership hiding is to formulate a function $h_{\params}(\cdot)$, parameterized by $\params$, that takes as input the initial graph $\mathcal{G}$ and produces as output a \textit{new} graph $h_{\params}(\G) = \mathcal{G'} = (\mathcal{V}, \mathcal{E'})$. Among all the possible graphs, we seek the one which, when input to the community detection algorithm $f$, disassociates a target node $u$ from its original community $\mathcal{C}_i$. 
This might lead to formulating various objectives depending on how we define community membership hiding. 
Let us consider the scenario where the target node $u$ is associated with a new community $\mathcal{C}'_i \in f(\G')$. 
One possible way to characterize community membership hiding for $u$ is to aim for a small similarity between the new community and the original community to which $u$ belonged.
Alternatively, one may opt to enforce specific changes, ensuring that the target node $u$ no longer shares the same community with \textit{some} nodes. For example, if $\mathcal{C}_i = \{s, t, \underline{u}, v, w, x, y, z\}$, we might wish to assign $u$ to a new community $\mathcal{C}'_i$ such that $s, w, z \notin \mathcal{C}'_i$.

In this work, we adopt the first definition, leaving exploration of other possibilities for future research. Practically, we set a similarity threshold between $\mathcal{C}'_i$ and $\mathcal{C}_i$, excluding the target node $u$, which belongs to both communities by definition. This condition can be expressed as $sim(\mathcal{C}_i - {u}, \mathcal{C}'_i - {u}) \leq \tau$, where $\tau \in [0,1]$. (Note: We assume that $sim(\cdot, \cdot)$ ranges between $0$ and $1$.)
Several similarity measures can be used to measure $sim(\cdot, \cdot)$ depending on the application domain, e.g., the overlap coefficient (a.k.a. Szymkiewicz–Simpson coefficient) \cite{metrics:Szymkiewicz_Simpson}, the Jaccard coefficient \cite{metrics:Jaccard}, and the S{\o}rensen-Dice coefficient \cite{metrics:Dice}.
Setting $\tau = 0$ represents the most stringent scenario, where we require zero overlaps between $\mathcal{C}'_i$ and $\mathcal{C}_i$, except for the node $u$ itself. At the other extreme, when $\tau = 1$, we adopt a more tolerant strategy, allowing for maximum overlap between $\mathcal{C}'_i$ and $\mathcal{C}_i$. However, note that except for the overlap coefficient, which can yield a value of $1$ even if one community is a subset of the other, the Jaccard and S{\o}rensen-Dice coefficients yield a value of $1$ only when the two communities are identical. 
In practice, setting $\tau = 1$ may lead to the undesired outcome of $\mathcal{C}'_i$ being equal to $\mathcal{C}_i$, thus contradicting the primary goal of community membership hiding. Therefore, it is common to let $\tau \in [0, 1)$ to avoid this scenario.

Moreover, it is essential to emphasize that executing $f$ on $\G'$ instead of the original $\G$ could potentially influence $(i)$ the community affiliations of nodes beyond the selected target, $u$, and $(ii)$ the eventual count of recognized communities (i.e., $|f(\G')| = k' \neq k = |f(\G)|$), providing that $f$ does not need this number fixed apriori as one of its inputs. 
Thus, community membership hiding must strike a balance between two conflicting goals. On the one hand, the target node $u$ must be successfully elided from the original community $\mathcal{C}_i$; on the other hand, the cost of such an operation -- i.e., the "distance" between $\G$ and $\G'$, and between $f(\G)$ and $f(\G')$ -- must be as small as possible.


Overall, we can define the following loss function associated with the community membership hiding task:
\begin{equation}
\label{eq:loss}
\Loss(h_{\params};\G,f,u) = \loss_{\text{decept}}(\G, h_{\params}(\G); f, u) + \lambda \loss_{\text{dist}}(\G, h_{\params}(\G);f).
\end{equation}

The first component ($\loss_{\text{decept}}$) penalizes when the goal is {\em not} satisfied. 
Let $\Gamma$ be the set of input graphs which do {\em not} meet the membership hiding objective, i.e., those which retain node $u$ as part of the community $\mathcal{C}_i$. 
More formally, let $\mathcal{\widetilde{C}}_i$ be the community to which node $u$ is assigned when $f$ is applied to the input graph $\widetilde{\G}$. We define $\Gamma = \{\widetilde{\G}~|~sim(\mathcal{C}_i - \{u\}, \mathcal{\widetilde{C}}_i - \{u\}) > \tau \}$.
Thus, we can compute $\loss_{\text{decept}}$ as follows:
\begin{equation}
\label{eq:ldecept}
\loss_{\text{decept}}(\G, h_{\params}(\G); f) = \mathbbm{1}_{\Gamma}(h_{\params}(\G)),
\end{equation}
where $\mathbbm{1}_{\Gamma}(h_{\params}(\G))$ is the well-known $0$-$1$ indicator function, which evaluates to $1$ if $h_{\params}(\G) \in \Gamma$, or $0$ otherwise. 


The second component, denoted as $\loss_{\text{dist}}$, is a composite function designed to assess the overall dissimilarity between two graphs and their respective communities found by $f$. This function serves the dual purpose of $(i)$ discouraging the new graph $h_{\params}(\G)$ from diverging significantly from the original graph $\G$ and $(ii)$ preventing the new community structure $f(h_{\params}(\G))$ from differing substantially from the prior community structure $f(\G)$.

\subsection{Counterfactual Graph Objective}
\label{subsec:cf-graph}
Given the target community $\mathcal{C}_i$, from which we want to exclude node $u$, we can classify the remaining nodes $\V - \{u\}$ of $\G$ into two categories: nodes that are inside the same community $\mathcal{C}_i$ as $u$ and nodes that belong to a different community from $u$.
%
%
This categorization helps us define which edges the target node $u$ can control and, thus, directly manipulate under the assumption that $\G$ is a directed graph.\footnote{We can easily extend this reasoning if $\G$ is \textit{undirected}.}
Specifically, following \cite{deception_modularity_3}, we assume that $u$ can $(i)$ \textit{remove} existing outgoing edges to nodes that are inside $u$'s community $(ii)$ \textit{add} new outgoing edges to nodes that are outside $u$'s community.
We intentionally exclude two possible actions: $(iii)$ removing outgoing links to outside-community nodes and $(iv)$ adding outgoing links to inside-community nodes.
Two primary reasons drive this choice. On the one hand, allowing $(iii)$ could isolate $u$ and its original community $\mathcal{C}_i$ further. On the other hand, allowing $(iv)$ would enhance connectivity between $u$ and other nodes in $\mathcal{C}_i$. 
Both events contradict the goal of community membership hiding.

Overall, we can define the set of candidate edges to remove ($\edges_{u,i}^-$) and to add ($\edges_{u,i}^+$) as follows:
\begin{align*} 
    \edges_{u,i}^- & = \{(u,v)~|~u,v\in \mathcal{C}_i \wedge (u,v)\in \edges\}, \\
    \edges_{u,i}^+ & = \{(u,v)~|~u \in \mathcal{C}_i, v\notin \mathcal{C}_i \wedge (u,v) \notin \edges \}.
\end{align*}



If we suppose the target node $u$ has a fixed budget $\beta > 0$, solving the community membership hiding task resorts to finding the optimal model $h^* = h_{\params^*}$ as the one whose parameters $\params^*$ minimize Eq.~(\ref{eq:loss}), i.e., by solving the following constrained objective: 
\begin{equation}
\label{eq:objective}
    \begin{split} 
    \params^* & = \underset{\params}{\text{arg min}} \bigg \lbrace \Loss(h_{\params};\G, f, u) \bigg \rbrace \\
    & \text{subject to: } |\mathcal{B}_{u,i}| \leq \beta,
    \end{split}
\end{equation}
where $\mathcal{B}_{u,i} \subseteq \edges_{u,i}^- \cup \edges_{u,i}^+$ is the set of graph edge modifications selected from the candidates.

Note that Eq.~(\ref{eq:objective}) resembles the optimization task to find the best \textit{counterfactual graph} $\G^* = h^*(\G)$ that, when fed back into $f$, changes its output to hide the target node $u$ from its community.

\subsection{Markov Decision Process}
The community membership hiding problem defined in Eq.~(\ref{eq:objective}) requires minimizing a discrete, non-differentiable loss function. 
Thus, standard optimization methods like stochastic gradient descent are unsuitable for this task. 
One potential solution is to smooth the loss function using numerical techniques, such as applying a real-valued perturbation matrix to the original graph's adjacency matrix like in \cite{lucic2022aistats,trappolini2023savage}. 
Another option is to directly define $\loss_{\text{decept}}$ as a smooth similarity between the original and the newly obtained community, sacrificing control over the threshold $\tau$.
We leave the exploration of these alternatives for future work.
Instead, we take a different approach and cast this problem as a sequential decision-making process, following standard reinforcement learning principles.

In this framework, at each time step, an agent: \emph{(i)} takes an action (choosing to add or remove an edge based on the rules defined above), and \emph{(ii)} observes the new set of communities output by $f$ when this is fed with the graph modified according to the action taken before. The agent also receives a scalar reward from the environment.
The process continues until the agent eventually meets the specified node hiding-goal and the optimal counterfactual graph $\G^*$ -- i.e., the optimal $h^*$ -- is found. 

We formalize this scheme as a discounted Markov decision process (MDP) denoted as $\mathcal{M} = \{\mathcal{S}, \mathcal{A}, \mathcal{P}, p_0, r, \gamma\}$ and detailed below.

\noindent \textbf{\textit{States} ($\mathcal{S}$).}
At each time step $t$, the agent's state is $S_t=s_t$, where $s_t = \G^t \in \mathcal{S}$ is the current modified input graph. 
In practice, though, we can replace $\G^t$ with its associated adjacency matrix $\bm{A}_t \in \{0,1\}^{n \times n}$.
Initially, when $t=0$, $\G^0 = \G$ ($\bm{A}^0 = \bm{A}$).

\noindent \textbf{\textit{Actions ($\mathcal{A}$).}}
The set of actions is defined by $\mathcal{A} = \{a_t\}$, which consists of all valid graph rewiring operations, assuming node $u$ belongs to the community $C_i$.
\begin{equation}
    \mathcal{A} = \{\text{del}(u, v) | (u,v) \in \edges_{u,i}^-\} \cup \{\text{add}(u, v) | (u,v) \in \edges_{u,i}^+ \}.
\end{equation}
According to the allowed graph modifications outlined in Section~\ref{subsec:cf-graph}, the agent can choose between two types of actions: deleting an edge from $u$ to any node within the same community $C_i$ or adding an edge from $u$ to any node in a different community.

\noindent \textbf{\textit{Transitions Probability} ($\mathcal{P}$).}
Let $a_t \in \mathcal{A}$ be the action taken by the agent at iteration $t$. This action deterministically guides the agent's transition from the state $s_t$ to the state $s_{t+1}$. In essence, the transition function $\mathcal{P}: \mathcal{S} \times \mathcal{A} \times \mathcal{S} \to [0,1]$, which associates a transition probability with each state-action pair, assigns a transition probability of $1$ when the subsequent state $s_{t+1}$ is determined by the state-action pair $(s_t, a_t)$ and $0$ otherwise. Formally:
\begin{itemize}
    \item $\mathcal{P}(s_{t+1} \vert s_t, a_t) = 1$ if $s_{t+1}$ is the next state resulting from the application of action $a_t$ in state $s_t$.
    \item $\mathcal{P}(s_{t+1} \vert s_t, a_t) = 0$ otherwise.
\end{itemize}

\noindent \textbf{\textit{Reward} ($r$).}
\label{sec:reward}
The reward function of the action $a_t$ which takes the agent from state $s_t$ to state $s_{t+1}$ can be defined as:
\begin{equation} 
\label{eq:xi}
r(s_t, a_t) = \begin{cases} 1 - \lambda (\loss_{\text{dist}}^t - \loss_{\text{dist}}^{t-1}) &\text{, if "the goal is met"} 
\\ - \lambda (\loss_{\text{dist}}^t - \loss_{\text{dist}}^{t-1}) &\text{, otherwise.} \end{cases}
\end{equation}
The goal is considered successfully achieved when $f(\G^t)$ leads to $u\in \mathcal{C}_i^t \neq \mathcal{C}_i$ such that $sim(\mathcal{C}_i - \{u\}, \mathcal{C}_i^t - \{u\}) \leq \tau$. 
In addition, $\loss_{\text{dist}}^t = \loss_{\text{dist}} (\G, \G^t;f)$ measures the penalty computed on the graph before and after action $a_t$, and $\lambda \in \mathbb{R}_{>0}$ is a parameter that controls its weight.
More precisely, the penalty is calculated as follows:
\begin{equation}
\label{eq:penalty}
    \loss_{\text{dist}} (\G, \G^t; f) = \alpha \times d_{\text{community}} + (1 - \alpha) \times d_{\text{graph}},
\end{equation}
where $d_{\text{community}}$ computes the distance between the community structures $f(\G)$ and $f(\G^t)$, $d_{\text{graph}}$ measures the distance between the two graphs $\G$ and $\G^t$, and the parameter $\alpha \in [0,1]$ balances the importance between the two distances.

Hence, the reward function encourages the agent to take actions that preserve the similarity between the community structures and the graphs before and after the rewiring action.

\noindent \textbf{\textit{Policy} ($\pi_{\params}$).}
We first define a parameterized policy $\pi_{\params}$ that maps from states to actions. We then want to find the values of the policy parameters $\params$ that maximize the expected reward in the MDP. This is equivalent to finding the optimal policy $\pi^*$, which is the policy that gives the highest expected reward for any state. We can find the optimal policy $\pi^*$ by minimizing the Eq.~($\ref{eq:objective}$). This minimization leads to the discovery of the optimal model $h^*$, which is the model that best predicts the rewards in the MDP.
\begin{equation}
\begin{split}
  \params^* & = \underset{\params}{\text{arg min }} \loss_{\text{decept}}(h_{\params}; \G, f, u) + \lambda \loss_{\text{dist}}(\G, h_{\params}(\G);f) \\
    & = \underset{\params}{\text{arg max }} - \loss_{\text{decept}}(h_{\params}; \G, f, u) - \lambda \loss_{\text{dist}}(\G, h_{\params}(\G);f) \\
    & = \underset{\params}{\text{arg max }} \sum_{t=1}^T\begin{cases} 1 - \lambda (\loss_{\text{dist}}^t - \loss_{\text{dist}}^{t-1}) &\text{, if "the goal is met" } 
    \\ - \lambda (\loss_{\text{dist}}^t - \loss_{\text{dist}}^{t-1}) &\text{, otherwise } \end{cases} \\
    & = \underset{\params}{\text{arg max }} \sum_{t=1}^T r(s_t, \pi_{\params}(s_t)),  
\end{split}   
\end{equation}
where $T$ is the maximum number of steps per episode taken by the agent and is therefore always less than the allowed number of graph manipulations, i.e., $T \leq \beta$.

\section{Proposed Method}
\label{sec:method}

To learn the optimal policy for our agent defined above, we use the \textit{Advantage Actor-Critic} (A2C) algorithm \cite{mnih2016asynchronous}, a popular deep reinforcement learning technique that combines the advantages of both policy-based and value-based methods. 
Specifically, A2C defines two neural networks, one for the policy function ($\pi_{\params}$) and another for the value function estimator ($V_v$), such that: 
\begin{align*}
    \nabla_{\params} \mathcal{J}(\params)  & \sim \underset{t=0}{\overset{T-1}{\sum}} \nabla_{\params} \log \pi_{\params} (a_t \vert s_t) A(s_t, a_t), \\
    \text{with } A(s_t, a_t) & = r_{t+1} + \gamma V_v(s_{t+1}) - V_v (s_t), \numberthis \label{eq:a2c}
\end{align*}
where $\mathcal{J}(\params)$ is the reward function, and the goal is to find the optimal policy parameters $\params$ that maximize it. $A(s_t, a_t)$ is the advantage function that quantifies how good or bad an action $a_t$ is compared to the expected value of actions chosen based on the current policy.

Below, we describe the policy (\textit{actor}) and value function (\textit{critic}) networks. 

\noindent \textbf{\textit{Policy Function Network (Actor)}.}
The policy function network is responsible for generating a probability distribution over possible actions based on the input, which consists of a list of nodes and the graph's feature matrix. 
However, some graphs may lack node features. In such cases, we can extract continuous node feature vectors (i.e., node embeddings) with graph representational learning frameworks like \textit{node2vec} \cite{node2vec}. These node embeddings serve as the feature matrix.

Our neural network implementation comprises a primary graph convolution layer (GCNConv \cite{gcnconv}) for updating node features. The output of this layer, along with skip connections, feeds into a block consisting of three hidden layers. Each hidden layer includes multi-layer perception (MLP) layers, ReLU activations, and dropout layers. The final output is aggregated using a sum-pooling function. In building our network architecture, we were inspired, in part, by the work conducted by \citet{network_architecture}, adapting it to our task.
The policy is trained to predict the probability that node $v$ is the optimal choice for adding or removing the edge $(u, v)$ to hide the target node $u$ from its original community. 
The feasible actions depend on the input node $u$ and are restricted to a subset of the graph's edges as outlined in Section~\ref{subsec:cf-graph}. Hence, not all nodes $v \in \mathcal{V}$ are viable options for the policy.

\noindent \textbf{\textit{Value Function Network (Critic)}.}
This network is similar to the one used for the policy, with one distinction: it includes a global sum-pooling operation on the convolution layer's output. This pooling operation results in an output layer with a size of 1, indicating the estimated value of the value function. The role of the value function is to predict the state value given a specific action $a_t$ and state $s_t$.

\section{Experiments}
\label{sec:experiments}

\subsection{Experimental Setup}
\label{subsec:exp-setup}

\noindent \textbf{\textit{Datasets.}} We train our DRL agent on the real dataset \texttt{words}.\footnote{\href{http://konect.cc/}{http://konect.cc/}\label{fn:dataset1}}
This dataset strikes a favorable balance regarding the number of nodes, edges, and discovered communities.
In addition, we evaluate the performance of our method on four additional datasets: \texttt{kar}\footref{fn:dataset1}, Wikipedia's \texttt{vote} \footnote{\href{https://networkrepository.com}{https://networkrepository.com}\label{fn:dataset2}}, \texttt{pow}\footref{fn:dataset1}, and Facebook \texttt{fb-75}\footref{fn:dataset2}.

\begin{table}[htb!]
\begin{center}
\caption{\label{tab:datasets_and_communities}Properties of the datasets used along with the size of the communities detected by \texttt{greedy}, \texttt{louvain}, and \texttt{walktrap}.}
\scalebox{0.9}{
\begin{tabular}{ | l | c | c || c | c | c | }
    \hline
    \multirow{2}{*} {\textbf{Dataset}} & \multirow{2}{*}{$|\mathcal{V}|$} & \multirow{2}{*}{$|\mathcal{E}|$} &  \multicolumn{3}{c|}{\textbf{Community Detection Algorithm}}\\
    \cline{4-6}
    {} & {} & {} & \texttt{greedy} & \texttt{louvain} & \texttt{walktrap} \\
    \hline
    \hline
    \texttt{kar} & 34 & 78 & 3 & 4 & 5\\
    \hline
    \texttt{words} & 112 & 425 & 7 & 7 & 25 \\
    \hline
    \texttt{vote} & 889 & 2,900 & 12 & 10 & 42 \\
    \hline
    \texttt{pow} & 4,941 & 6,594 & 40 & 41 & 364 \\
    \hline
    \texttt{fb-75} & 6,386 & 217,662 & 29 & 19 & 357 \\
    \hline
\end{tabular}
}
\end{center}
\end{table}

\noindent \textbf{\textit{Community Detection Algorithms.}} The DRL agent is trained using a single detection algorithm, namely the modularity-based Greedy (\texttt{greedy}) algorithm \cite{greedy_detection_alg}. However, at test time, we use two additional algorithms: the Louvain (\texttt{louvain}) algorithm \cite{louvain_detection_alg}, which is in the same family as \texttt{greedy}, and WalkTrap (\texttt{walktrap}) \cite{walktrap_detection_alg}, which takes a distinct approach centered on Random Walks.

Table~\ref{tab:datasets_and_communities} provides an overview of the datasets used, including key properties and the number of communities detected by each community detection algorithm. The modularity-based algorithms (\texttt{greedy} and \texttt{louvain}) yield comparable community counts, while \texttt{walktrap} identifies a generally higher number of communities. 


\noindent \textbf{\textit{Similarity/Distance Metrics.}}
To assess the achievement of the community membership hiding goal, i.e., whether the new community $\mathcal{C}_i^t$ of the target node $u$ at step $t$ can no longer be considered the same as the initial community $\mathcal{C}_i$, we need to define the $sim(\cdot)$ function used within $\loss_{\text{decept}}$. 
We employ the S{\o}rensen-Dice coefficient \cite{metrics:Dice}, which is defined as follows:
\begin{equation}
\label{eq:sdc}
    \text{DSC}(\mathcal{C}_i, \mathcal{C}_i^t) = \frac{2 \vert \mathcal{C}_i \bigcap \mathcal{C}_i^t \vert}{\vert \mathcal{C}_i \vert + \vert \mathcal{C}_i^t \vert}.
\end{equation}
This metric returns a value between $0$ (no similarity) and $1$ (strong similarity). If that value is less than or equal to the parameter $\tau$, we consider the node-hiding goal successfully met.

Furthermore, as described in Eq.~(\ref{eq:penalty}), we must specify the penalty function ($\loss_{\text{dist}}$), which consists of two mutually balanced factors, namely $d_{\text{community}}$ and $d_{\text{graph}}$. These factors quantify the dissimilarity between community structures and graphs before and after the action. 
During model training, we operationalize these distances using the Normalized Mutual Information (NMI) score for community comparison and the Jaccard distance for graph comparison.

The NMI score \cite{nmi_1, nmi_2}, utilized for measuring the similarity between community structures, ranges from $0$ (indicating no mutual information) to $1$ (indicating perfect correlation). 

Following the formulation by \cite{nmi_3}, it can be expressed as follows:
\begin{equation}
\label{eq:NMI}
    \text{NMI} ( \mathcal{K}, \mathcal{K}^t ) = I_{\text{norm}}  (X:Y) = \frac{H(X) + H(Y) - H(X, Y)}{(H(X) + H(Y)) / 2},
\end{equation}
where $H(X)$ and $H(Y)$ denote the entropy of the random variables $X$ and $Y$ associated with partitions $\mathcal{K} = f(\G)$ and $\mathcal{K}^t = f(\G^t)$, respectively, while $H(X, Y)$ denotes the joint entropy. 
Since we need to transform this metric into a distance, we calculate $1$-NMI.

The Jaccard distance can be adapted to the case of two graphs, as described in \cite{donnat2018tracking}, as follows:
\begin{equation}
    \text{Jaccard}(\G, \G^t) = 
    \frac{\vert \G \bigcup \G^t \vert  - \vert \G \bigcap \G^t \vert}{|\G \bigcup \G^t|} = \frac{\sum_{i,j} \vert \bm{A}_{i,j} - \bm{A}_{i,j}^{t} \vert}{\sum_{i,j \text{ max} (\bm{A}_{i,j}, \bm{A}_{i,j}^{t}) }},
\end{equation}
where $\bm{A}_{i,j}$ denotes the $(i,j)$-th entry of the adjacency matrix for the original graph $\G$. 
The Jaccard distance yields a value of $0$ when the two graphs are identical and $1$ when they are entirely dissimilar.


\subsection{Community Membership Hiding Task}
\label{subsec:node-dec}
\noindent \textbf{\textit{Baselines.}}
We compare the performance of our method (\textit{DRL-Agent})  against the five baselines described below.

\noindent \textit{1) Random-based.} This baseline operates by randomly selecting one of the nodes in the graph. If the selected node is a neighbor of the target node to be hidden (i.e., there is an edge between them), the edge is removed; otherwise, it is added. The randomness of these decisions aims to obscure the node's true community membership.

\noindent \textit{2) Degree-based.} This approach selects nodes with the highest degrees within the graph and rewires them. By prioritizing nodes with higher degrees, this baseline seeks to disrupt the node's central connections within its initial community, thus promoting concealment.

\noindent \textit{3) Betweenness-based.} This baseline prioritizes nodes with the highest \textit{betweenness centrality}~\cite{freeman1977betweenness} as those candidates to be disconnected. The betweenness centrality of a node $u$, denoted as $b(u)$, is computed as $b(u) = \sum_{s\neq u \neq t}\frac{\sigma_{s,t}(u)}{\sigma_{s,t}}$, where $\sigma_{s,t}$ is the total number of shortest paths from node $s$ to node 
$t$ and $\sigma_{s,t}(u)$ is the number of those paths that pass through 
$u$ (where $u$ is \textit{not} an end point).

\noindent \textit{4) Roam-based.} This method is based on the Roam heuristics \cite{deception_modularity_2}, originally designed to reduce a node's centrality within the network. It aims to diminish the centrality and influence of the target node within its initial community, favoring its deception.

\noindent \textit{5) Greedy-based.} 
We develop a "relaxed" greedy heuristic that selects the most promising rewiring action without exhaustively exploring all the possible operations by prioritizing higher-degree nodes.

Let $u$ be the target node to be masked from the community $\mathcal{C}$. At each step, we still can choose between $(i)$ adding an edge between $u$ and a node $v$ outside the community or $(ii)$ deleting an existing edge between $u$ and a node $w$ inside the same community. To greedily determine what action provokes the largest loss reduction as of Eq. (\ref{eq:loss}), we restrict the two actions above as follows.

Concerning $(i)$, we do not consider all the possible nodes $v \in \mathcal{V} \setminus \mathcal{C}$ but only the node $v^*$ with the highest out-degree. 
Then, we measure the loss reduction once the edge $(u,v^*)$ is added. The rationale is similar to the one already used in the Degree-based baseline, where the goal is to connect $u$ to a popular node and, hopefully, mitigate its "centrality" in its current community $\mathcal{C}$.
Concerning $(ii)$, we select the edge to remove as follows. For each node $w\neq u$ that belongs to the same community $\mathcal{C}$, such that $(u,w) \in \mathcal{E}$, we calculate the intra-community degree $deg_{\mathcal{C}}(w) = |\{(w,x) \in \mathcal{E}|x \in \mathcal{C}\}|$. 
Then, we delete the edge $(u,w^*)$ with the node $w^*$ having the highest intra-community degree. 
The rationale of this choice is to remove the link between $u$ and the node with the largest number of connections within the community. By doing so, we expect to increase the chance for $u$ to be masked from $\mathcal{C}$. Again, we measure the loss reduction if the edge $(u,w^*)$ is deleted.

At each step, we decide between adding $(u,v^*)$ or deleting $(u,w^*)$ by selecting the action that results in the highest loss reduction. This relaxed greedy approach is computationally more tractable than the full-greedy heuristic, as it allows us to run the community detection algorithm $f(\cdot)$ only twice at each step, once for each of the two actions. We keep doing this until the goal is achieved or the budget $\beta$ is exhausted.


\noindent \textbf{\textit{Evaluation Metrics.}} We measure the performance of each method in solving the community membership hiding task using the following metrics.

\noindent \textit{1) Success Rate} (SR). This metric calculates the success rate of the membership hiding algorithm by determining the percentage of times the target node is successfully hidden from its original community. If the target node no longer belongs to the original community (as per Eq.~(\ref{eq:sdc}) and the $\tau$ constraint), we consider the goal achieved. By repeating this procedure for several nodes and communities, we can estimate the algorithm's success rate. A higher value of this metric indicates better performance.

\noindent \textit{2) Normalized Mutual Information} (NMI). To quantify the impact of the function $h_{\params}(\cdot)$ on the resulting community structure, denoted as the output of $f(\G')$ where $\G'$ is the graph created by modifying the original graph $\G$, we compute NMI($\mathcal{K}, \mathcal{K}'$), as outlined in Eq.~(\ref{eq:NMI}). 
This score measures the similarity between the two structures, $\mathcal{K}=f(\G)$ and $\mathcal{K}'=f(\G')$. 
A higher value for this metric indicates a greater degree of similarity between the original and modified community structures and, therefore, a smaller cost.

In general, SR and NMI are two contrastive metrics; a higher SR corresponds to a lower NMI and vice versa. 
Therefore, similar to the $F1$ score, which balances precision and recall, we calculate the harmonic mean between SR and NMI using the formula $\frac{2 \times \text{SR} \times \text{NMI}}{\text{SR} + \text{NMI}}$ to evaluate which method achieves the optimal trade-off.


\subsection{Results and Discussion}
\label{subsec:results}

We assess our \textit{DRL-Agent}'s performance against the baseline methods discussed in Section~\ref{subsec:node-dec}. 
We investigate various parameter settings, including different values for the similarity constraint $\tau$ (${0.3, 0.5, 0.8}$) and the budget $\beta$ (${\frac{1}{2} \mu, 1 \mu, 2 \mu}$, where $\mu = \frac{|\edges|}{|\mathcal{V}|}$ ). 
We evaluate all possible combinations of these parameters.

For each parameter combination, dataset, and community detection algorithm, we conducted a total of 100 experiments. In each iteration, we randomly select a node from a different community than the previous one for concealment. The reported results are based on the average outcomes across all runs.
Furthermore, we explore two distinct setups: \textit{symmetric} and \textit{asymmetric}.
In the symmetric case, our \textit{DRL-Agent} is trained and tested using the \textit{same} community detection algorithm used for the membership hiding task.
Instead, the asymmetric setup evaluates the performance of our method when tested on a community detection algorithm different from the one used for training. 
This second setting allows us to assess the \textit{transferability} of our method to community detection algorithms unseen at training time.
Specifically, the results below showcase the outcome when we train our \textit{DRL-Agent} using the modularity-based \texttt{greedy} algorithm for both setups. In the symmetric setting, \texttt{greedy} is also used at test time, while in the asymmetric setup, we employ the \texttt{louvain} or \texttt{walktrap} algorithm.

All the experiments were conducted on a 4-core Intel Xeon CPU running at 2.2 GHz with 18 GB RAM. 

In Tables ~\ref{tab:sr_beta_sym} and ~\ref{tab:sr_beta_asym}, we display the Success Rate (SR) as a function of budget ($\beta$) for each dataset in the symmetric and two asymmetric settings, respectively, with a tolerance threshold of $\tau=0.5$.
\begin{table*}[htbp!]
    \caption{\label{tab:sr_beta_sym}Success Rate (SR) vs. budget ($\boldsymbol{\beta}$) for community membership hiding task in the \textit{symmetric} setting ($\boldsymbol{\tau=0.5}$).}
    \centering
    \scalebox{0.67}{
    \begin{tabular}{|c|c|c|c|c|c|c|c|}
        \hline
        \multirow{4}{*}{\textbf{Dataset}} & \multirow{4}{*}{$\bm{\beta}$} & \multicolumn{6}{c|}{\textbf{\textit{Symmetric} (training: \texttt{greedy}; testing: \texttt{greedy})}} 
        \\ 
        \cline{3-8} 
        & & \thead{DRL-Agent\\(ours)} & \textit{Random} & \textit{Degree} & \textit{Betweenness} & \textit{Roam} & \textit{Greedy}\\
        \hline
        \multirow{3}{*}{\texttt{kar}} & $\frac{1}{2}\mu$ & $\mathbf{26.3\% \pm 4.9\%}$ & $22.0\% \pm 4.6\%$ & $16.3\% \pm 4.1\%$ & $14.6\% \pm 4.0\%$ & $13.6\% \pm 3.8\%$ & $0.6\% \pm 0.9\%$ \\
        & $1\mu$ & $\mathbf{54.0\% \pm 5.6\%}$ & $31.6\% \pm 5.2\%$ & $42.3\% \pm 5.5\%$ & $46.0\% \pm 5.6\%$ & $12.6\% \pm 3.7\%$ & $45.3\% \pm 5.6\%$ \\
        & $2\mu$ & $\mathbf{70.3\% \pm 5.1\%}$ & $46.6\% \pm 5.6\%$ & $43.6\% \pm 5.6\%$ & $55.6\% \pm 5.6\%$ & $13.6\% \pm 3.8\%$ & $38.3\% \pm 5.5\%$ \\
        \hline
        \multirow{3}{*}{\texttt{words}} & $\frac{1}{2} \mu$ & $\mathbf{49.7\% \pm 5.7\%}$ & $48.7\% \pm 5.7\%$ & $48.0\% \pm 5.7\%$ & $46.0\% \pm 5.6\%$ & $45.3\% \pm 5.6\%$ & $29.0\% \pm 5.1\%$ \\
        & $1 \mu$ & $\mathbf{71.7\% \pm 5.1\%}$ & $57.0\% \pm 5.6\%$ & $62.3\% \pm 5.5\%$ & $64.6\% \pm 5.4\%$ & $51.7\% \pm 5.7\%$ & $64.0\% \pm 5.4\%$ \\
        & $2 \mu$ & $\mathbf{88.3\% \pm 3.6\%}$ & $71.0\% \pm 5.1\%$ & $77.3\% \pm 4.7\%$ & $79.6\% \pm 4.5\%$ & $44.7\% \pm 5.6\%$ & $79.7\% \pm 4.5\%$ \\
        \hline
        \multirow{3}{*}{\texttt{vote}} & $\frac{1}{2} \mu$ & $\mathbf{38.6\% \pm 5.5\%}$ & $16.6\% \pm 4.2\%$ & $20.6\% \pm 4.5\%$ & $21.0\% \pm 4.6\%$ & $26.0\% \pm 4.9\%$ & $3.3\% \pm 2.0\%$ \\
        & $1 \mu$ & $\mathbf{49.6\% \pm 5.6\%}$ & $35.3\% \pm 5.4\%$ & $37.3\% \pm 5.4\%$ & $38.6\% \pm 5.5\%$ & $22.3\% \pm 4.7\%$ & $38.3\% \pm 5.5\%$ \\
        & $2 \mu$ & $\mathbf{65.3\% \pm 5.3\%}$ & $47.0\% \pm 5.6\%$ & $53.0\% \pm 5.6\%$ & $55.0\% \pm 5.6\%$ & $40.0\% \pm 5.5\%$ & $53.3\% \pm 5.6\%$ \\
        \hline
       \multirow{3}{*}{\texttt{pow}} & $\frac{1}{2} \mu$ & $\mathbf{40.0\% \pm 5.5\%}$ & $33.3\% \pm 5.3\%$ & $19.3\% \pm 4.4\%$ & $20.3\% \pm 4.5\%$ & $32.3\% \pm 5.2\%$ & $6.7\% \pm 2.8\%$ \\
        & $1 \mu$ & $\mathbf{71.3\% \pm 5.1\%}$ & $38.0\% \pm 5.4\%$ & $15.0\% \pm 4.0\%$ & $64.0\% \pm 5.4\%$ & $38.6\% \pm 5.5\%$ & $28.7\% \pm 5.1\%$ \\
        & $2 \mu$ & $91.6\% \pm 3.1\%$ & $47.3\% \pm 5.6\%$ & $91.6\% \pm 3.1\%$ & $\mathbf{93.0\% \pm 2.8\%}$ & $28.6\% \pm 5.1\%$ & $92.0\% \pm 3.0\%$ \\
        \hline
        \multirow{3}{*}{\texttt{fb-75}} & $\frac{1}{2} \mu$ & $\mathbf{29.0\% \pm 5.1\%}$ & $20.3\% \pm 4.5\%$ & $8.3\% \pm 3.1\%$ & $10.3\% \pm 3.4\%$ & $6.3\% \pm 2.7\%$ & $11.3\% \pm 3.6\%$ \\
        & $1 \mu$ & $\mathbf{33.6\% \pm 5.3\%}$ & $24.3\% \pm 4.8\%$ & $5.6\% \pm 2.6\%$ & $9.6\% \pm 3.3\%$ & $5.3\% \pm 2.5\%$ & $10.3\% \pm 3.4\%$ \\
        & $2 \mu$ & $\mathbf{45.0\% \pm 5.6\%}$ & $36.3\% \pm 5.4\%$ & $11.0\% \pm 3.5\%$ & $16.3\% \pm 4.1\%$ & $8.0\% \pm 3.0\%$ & $17.3\% \pm 4.3\%$ \\
        \hline
    \end{tabular}
    }
\end{table*}

\begin{table*}[htbp!]
    \caption{\label{tab:sr_beta_asym}Success Rate (SR) vs. budget ($\boldsymbol{\beta}$) community membership hiding task in two \textit{asymmetric} settings ($\boldsymbol{\tau=0.5}$).}
    \centering
    \scalebox{0.67}{
    \begin{tabular}{|c|c|c|c|c|c|c|c||c|c|c|c|c|c|}
        \hline
        \multirow{4}{*}{\textbf{Dataset}} & \multirow{4}{*}{$\bm{\beta}$} & \multicolumn{6}{c||}{\textbf{\textit{Asymmetric} (training: \texttt{greedy}; testing: \texttt{louvain})}} 
        &
        \multicolumn{6}{c|}{\textbf{\textit{Asymmetric} (training: \texttt{greedy}; testing: \texttt{walktrap})}}
        \\ 
        \cline{3-14} 
        & & \thead{DRL-Agent\\(ours)} & \textit{Random} & \textit{Degree} & \textit{Betweenness} & \textit{Roam} & \textit{Greedy}
        &
        \thead{DRL-Agent\\(ours)} & \textit{Random} & \textit{Degree} & \textit{Betweenness} & \textit{Roam} & \textit{Greedy}
        \\
        \hline
        \multirow{3}{*}{\texttt{kar}} & $\frac{1}{2}\mu$ & $\mathbf{33.3\% \pm 5.3\%}$ & $31.3\% \pm 5.2\%$ & $24.0\% \pm 4.8\%$ & $7.6\% \pm 3.0\%$ & $27.0\% \pm 5.0\%$ & $1.0\% \pm 1.1\%$  & $\mathbf{10.0\% \pm 3.4\%}$ & $6.7\% \pm 2.8\%$ & $6.7\% \pm 2.8\%$ & $4.0\% \pm 2.2\%$ & $6.0\% \pm 2.7\%$ & $4.0\% \pm 2.2\%$ \\
        & $1\mu$ & $\mathbf{50.0\% \pm 5.6\%}$ & $41.0\% \pm 5.5\%$ & $39.6\% \pm 5.5\%$ & $25.0\% \pm 4.9\%$ & $28.3\% \pm 5.1\%$ & $25.0\% \pm 4.9\%$ & $\mathbf{41.7\% \pm 5.6\%}$ & $30.3\% \pm 5.2\%$ & $19.7\% \pm 4.5\%$ & $26.7\% \pm 5.0\%$ & $9.3\% \pm 3.3\%$ & $25.0\% \pm 4.9\%$ \\
        & $2\mu$ & $\mathbf{66.6\% \pm 5.3\%}$ & $46.6\% \pm 5.6\%$ & $41.0\% \pm 5.5\%$ & $37.0\% \pm 5.4\%$ & $24.6\% \pm 4.8\%$ & $46.0\% \pm 5.6\%$ & $\mathbf{88.3\% \pm 3.6\%}$ & $61.0\% \pm 5.5\%$ & $58.3\% \pm 5.6\%$ & $70.3\% \pm 5.2\%$ & $7.7\% \pm 3.0\%$ & $74.3\% \pm 4.9\%$ \\
        \hline
        \multirow{3}{*}{\texttt{words}} & $\frac{1}{2} \mu$ & $\mathbf{57.6\% \pm 5.5\%}$ & $44.0\% \pm 5.6\%$ & $48.6\% \pm 5.6\%$ & $52.0\% \pm 5.6\%$ & $50.6\% \pm 5.6\%$ & $50.0\% \pm 5.7\%$ & $\mathbf{43.0\% \pm 5.6\%}$ & $35.7\% \pm 5.4\%$ & $20.3\% \pm 4.5\%$ & $29.7\% \pm 5.2\%$ & $32.0\% \pm 5.3\%$ & $21.0\% \pm 4.6\%$ \\
        & $1 \mu$ & $\mathbf{68.3\% \pm 5.2\%}$ & $51.6\% \pm 5.6\%$ & $54.6\% \pm 5.6\%$ & $60.3\% \pm 5.5\%$ & $59.3\% \pm 5.5\%$ & $60.0\% \pm 5.5\%$ & $\mathbf{64.0\% \pm 5.4\%}$ & $52.7\% \pm 5.7\%$ & $41.0\% \pm 5.6\%$ & $31.3\% \pm 5.2\%$ & $26.0\% \pm 5.0\%$ & $43.7\% \pm 5.6\%$ \\
        & $2 \mu$ & $\mathbf{84.0\% \pm 4.1\%}$ & $61.6\% \pm 5.5\%$ & $59.6\% \pm 5.5\%$ & $69.0\% \pm 5.2\%$ & $56.3\% \pm 5.6\%$ & $83.0\% \pm 4.2\%$ & $\mathbf{76.7\% \pm 4.8\%}$ & $66.3\% \pm 5.3\%$ & $65.0\% \pm 5.4\%$ & $51.7\% \pm 5.7\%$ & $42.0\% \pm 5.6\%$ & $57.0\% \pm 5.6\%$ \\
        \hline
        \multirow{3}{*}{\texttt{vote}} & $\frac{1}{2} \mu$ & $\mathbf{22.0\% \pm 4.6\%}$ & $19.0\% \pm 4.4\%$ & $14.3\% \pm 3.9\%$ & $8.0\% \pm 3.0\%$ & $20.3\% \pm 4.5\%$ & $7.0\% \pm 2.9\%$ & $\mathbf{27.0\% \pm 5.0\%}$ & $23.0\% \pm 4.8\%$ & $13.3\% \pm 3.9\%$ & $12.7\% \pm 3.8\%$ & $25.0\% \pm 4.9\%$ & $7.0\% \pm 2.9\%$ \\
        & $1 \mu$ & $\mathbf{35.6\% \pm 5.4\%}$ & $28.6\% \pm 5.1\%$ & $23.0\% \pm 4.7\%$ & $16.3\% \pm 4.1\%$ & $29.6\% \pm 5.1\%$ & $26.7\% \pm 5.0\%$  & $\mathbf{45.7\% \pm 5.6\%}$ & $35.3\% \pm 5.4\%$ & $38.7\% \pm 5.5\%$ & $39.3\% \pm 5.5\%$ & $32.0\% \pm 5.3\%$ & $29.7\% \pm 5.2\%$ \\
        & $2 \mu$ & $\mathbf{45.0\% \pm 5.6\%}$ & $35.6\% \pm 5.4\%$ & $39.6\% \pm 5.5\%$ & $21.6\% \pm 4.6\%$ & $31.0\% \pm 5.2\%$ & $40.0\% \pm 5.5\%$ & $\mathbf{66.7\% \pm 5.3\%}$ & $51.3\% \pm 5.7\%$ & $64.0\% \pm 5.4\%$ & $65.0\% \pm 5.4\%$ & $42.7\% \pm 5.6\%$ & $64.7\% \pm 5.4\%$ \\
        \hline
       \multirow{3}{*}{\texttt{pow}} & $\frac{1}{2} \mu$ & $\mathbf{56.6\% \pm 5.6\%}$ & $41.3\% \pm 5.5\%$ & $19.3\% \pm 4.4\%$ & $15.0\% \pm 4.0\%$ & $45.3\% \pm 5.6\%$ & $11.3\% \pm 3.6\%$ & $49.7\% \pm 5.7\%$ & $37.7\% \pm 5.5\%$ & $29.0\% \pm 5.1\%$ & $5.3\% \pm 2.5\%$ & $\mathbf{65.3\% \pm 5.4\%}$ & $7.3\% \pm 2.9\%$ \\
        & $1 \mu$ & $\mathbf{63.3\% \pm 5.4\%}$ & $46.6\% \pm 5.6\%$ & $21.6\% \pm 4.6\%$ & $49.6\% \pm 5.6\%$ & $45.6\% \pm 5.6\%$ & $25.0\% \pm 4.9\%$ &  $56.0\% \pm 5.6\%$ & $42.7\% \pm 5.6\%$ & $33.0\% \pm 5.3\%$ & $8.0\% \pm 3.1\%$ & $\mathbf{64.0\% \pm 5.4\%}$ & $24.7\% \pm 4.9\%$ \\
        & $2 \mu$ & $82.3\% \pm 4.3\%$ & $55.3\% \pm 5.6\%$ & $32.3\% \pm 5.2\%$ & $\mathbf{96.0\% \pm 2.2\%}$ & $47.0\% \pm 5.6\%$ & $43.3\% \pm 5.6\%$ &  $\mathbf{62.0\% \pm 5.5\%}$ & $46.0\% \pm 5.6\%$ & $48.0\% \pm 5.7\%$ & $27.7\% \pm 5.1\%$ & $44.3\% \pm 5.6\%$ & $36.3\% \pm 5.4\%$ \\
        \hline
        \multirow{3}{*}{\texttt{fb-75}} & $\frac{1}{2} \mu$ & $\mathbf{48.3\% \pm 5.6\%}$ & $42.0\% \pm 5.5\%$ & $37.0\% \pm 5.4\%$ & $41.0\% \pm 5.5\%$ & $36.3\% \pm 5.4\%$ & $41.0\% \pm 5.6\%$ & $\mathbf{31.7\% \pm 5.3\%}$ & $20.3\% \pm 4.5\%$ & $10.3\% \pm 3.4\%$ & $12.0\% \pm 3.7\%$ & $12.0\% \pm 3.7\%$ & $14.3\% \pm 3.9\%$ \\
        & $1 \mu$ & $\mathbf{56.0\% \pm 5.6\%}$ & $48.6\% \pm 5.6\%$ & $41.6\% \pm 5.5\%$ & $43.3\% \pm 5.6\%$ & $34.3\% \pm 5.3\%$ & $42.3\% \pm 5.6\%$ & $\mathbf{32.3\% \pm 5.3\%}$ & $27.0\% \pm 5.0\%$ & $13.0\% \pm 3.8\%$ & $19.0\% \pm 4.4\%$ & $12.3\% \pm 3.7\%$ & $23.3\% \pm 4.8\%$ \\
        & $2 \mu$ & $\mathbf{66.3\% \pm 5.3\%}$ & $63.6\% \pm 5.4\%$ & $60.0\% \pm 5.5\%$ & $64.0\% \pm 5.4\%$ & $48.0\% \pm 5.6\%$ & $62.0\% \pm 5.5\%$ & $\mathbf{36.0\% \pm 5.4\%}$ & $33.7\% \pm 5.3\%$ & $16.3\% \pm 4.2\%$ & $22.3\% \pm 4.7\%$ & $18.3\% \pm 4.4\%$ & $25.7\% \pm 4.9\%$ \\
        \hline
    \end{tabular}
    }
\end{table*}
In Figures~\ref{fig:sr-nmi-hm_symmetric}, \ref{fig:sr-nmi-hm_asymmetric}, and \ref{fig:sr-nmi-hm_asymmetric_walktrap}, we show the harmonic mean ($F1$ score) between the Success Rate and NMI for all datasets in symmetric and asymmetric setups, still with a tolerance threshold of $\tau=0.5$ and a budget $\beta=1\mu$.
\begin{figure}[htbp!]
    \centering    
    \includegraphics[width=.95\columnwidth]{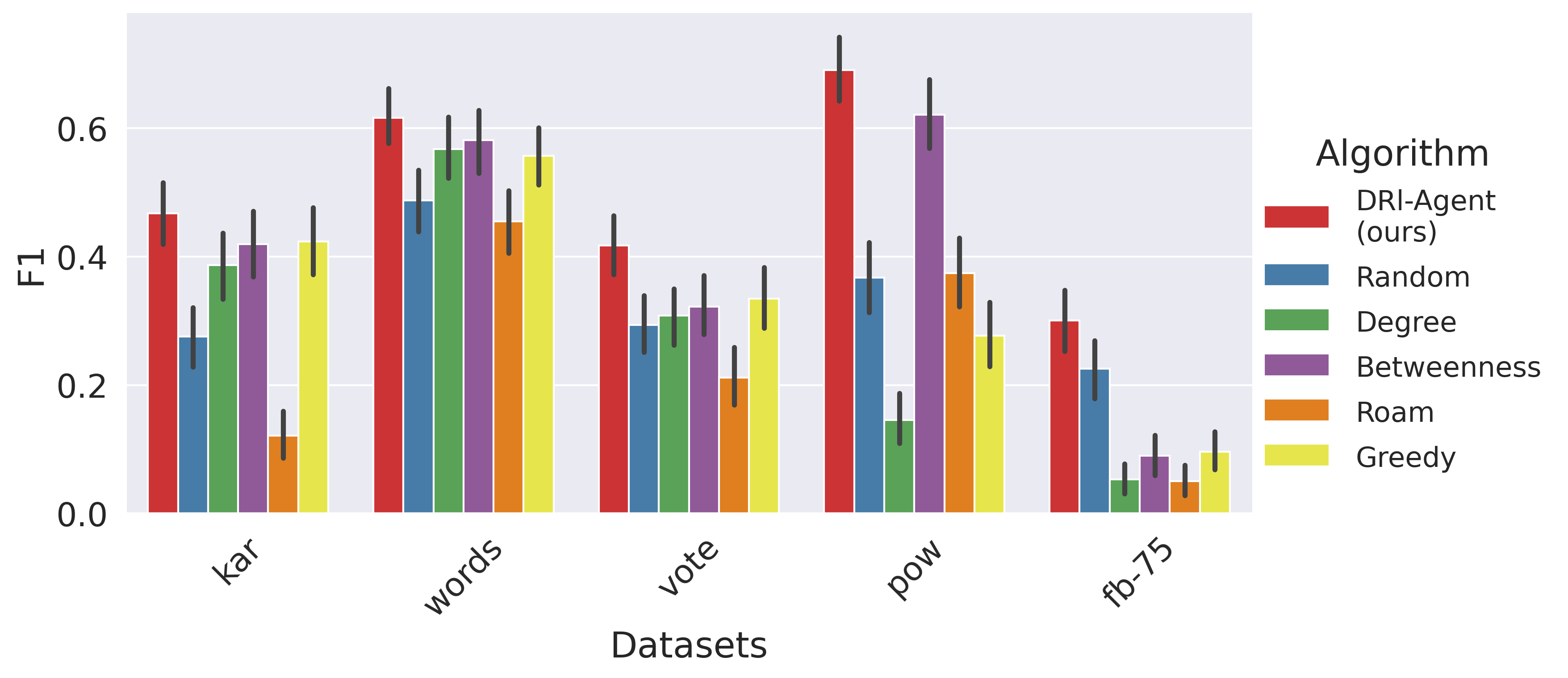}
    \caption{\textbf{$\mathbf{F1}$ score of SR and NMI in \textit{symmetric} setting (training: \texttt{greedy}; testing: \texttt{greedy}; $\boldsymbol{\tau=0.5}$; $\boldsymbol{\beta=1\mu}$).}}
    \label{fig:sr-nmi-hm_symmetric}
    \end{figure}
\begin{figure}[htbp!]
    \centering
     \includegraphics[width=.95\columnwidth]{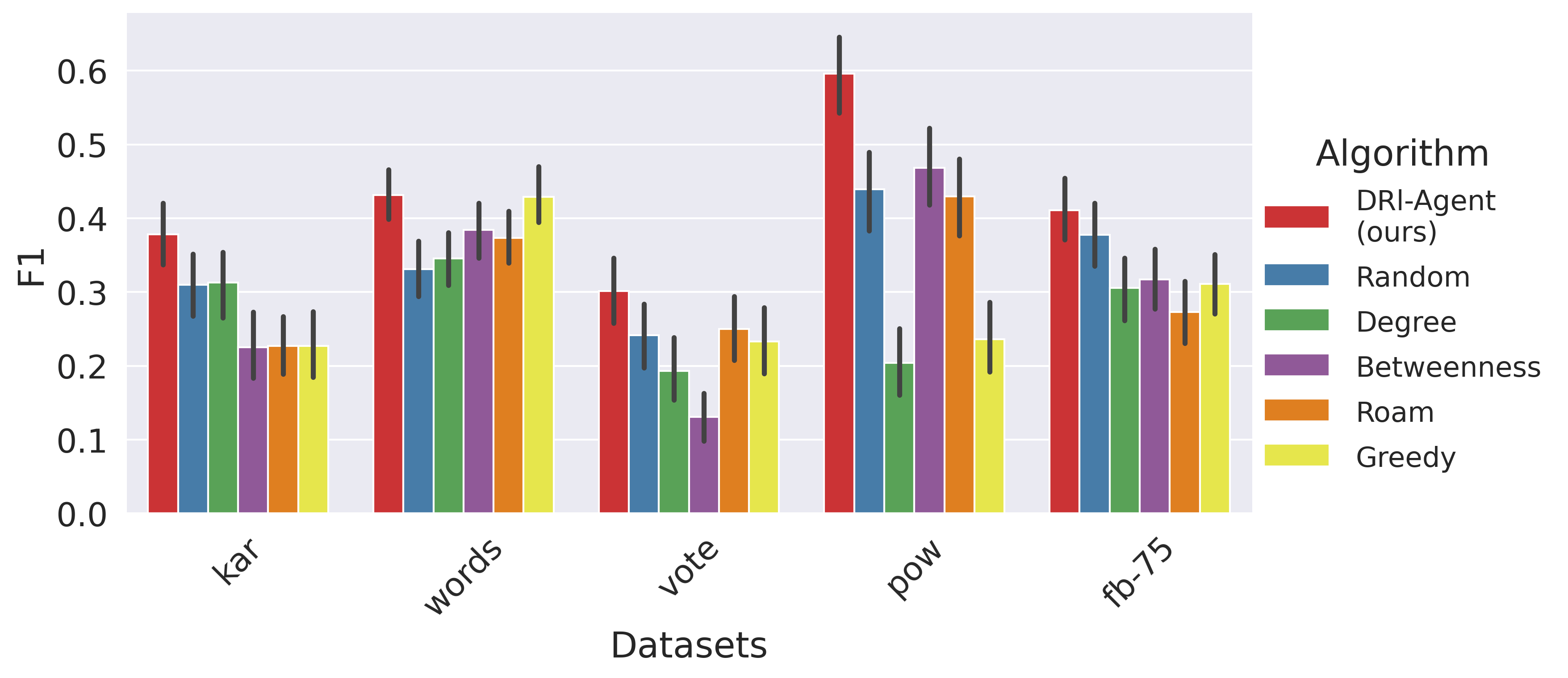}
     \caption{\textbf{$\mathbf{F1}$ score of SR and NMI in \textit{asymmetric} setting (training: \texttt{greedy}; testing: \texttt{louvain}; $\boldsymbol{\tau=0.5}$; $\boldsymbol{\beta=1\mu}$).}}
     \label{fig:sr-nmi-hm_asymmetric}
\end{figure}
\begin{figure}[htbp!]
    \centering
     \includegraphics[width=.95\columnwidth]{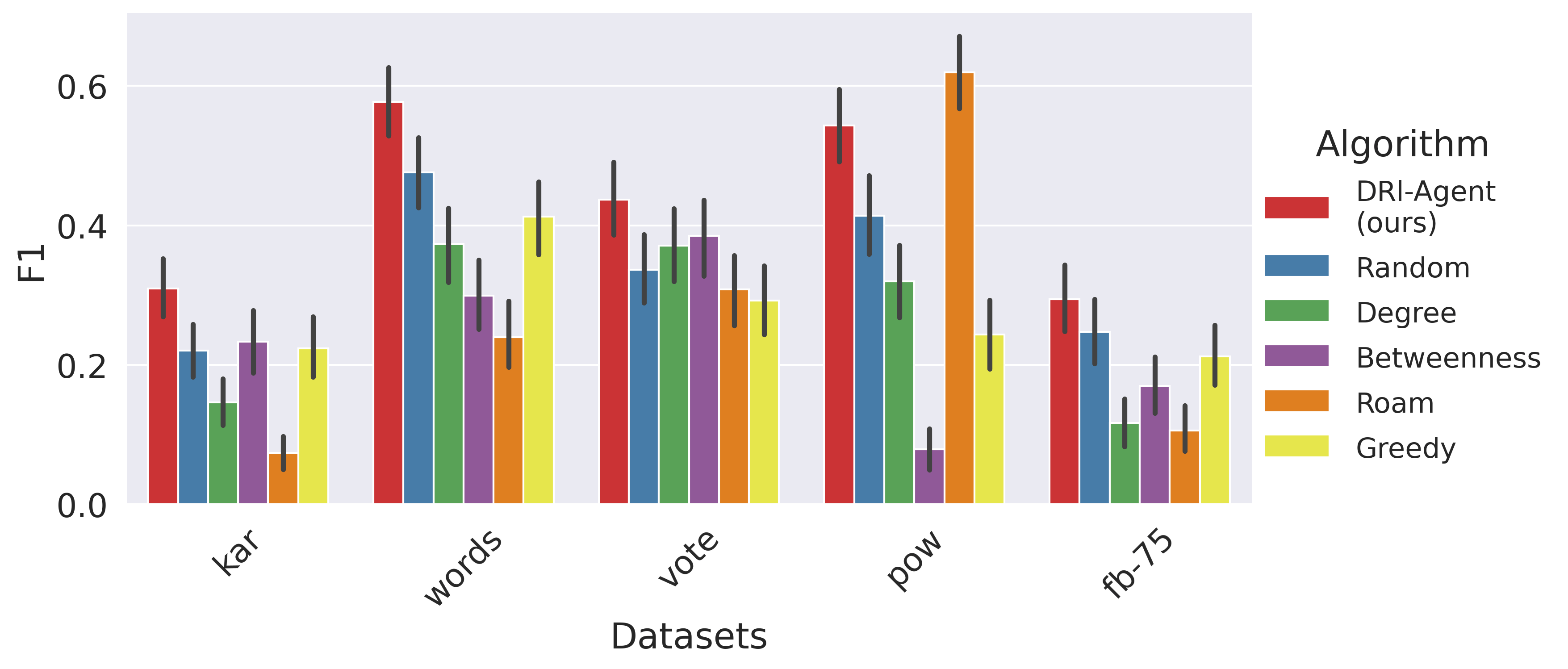}
     \caption{\textbf{$\mathbf{F1}$ score of SR and NMI in \textit{asymmetric} setting (training: \texttt{greedy}; testing: \texttt{walktrap}; $\boldsymbol{\tau=0.5}$; $\boldsymbol{\beta=1\mu}$).}}
     \label{fig:sr-nmi-hm_asymmetric_walktrap}
\end{figure}

From the results above, we emphasize two primary findings. 
Firstly, our approach strikes the best balance between accuracy and cost. 
Indeed, for a fixed budget value, our \textit{DRL-Agent} achieves the highest Success Rate in hiding the selected target node compared to other competing methods. The only exception occurs in a single dataset (\texttt{pow}) and for a specific budget value ($\beta=2\mu$), where the method based on \textit{Betweenness} outperforms all others. 
Moreover, our approach pays a limited "price" for this success. Specifically, the modified graph retains much of the structural properties of the original. This is demonstrated by the highest $F1$ score between SR and NMI compared to other methods.

The second key finding concerns the transferability of our \textit{DRL-Agent} to a different community detection algorithm than the one seen during training (\textit{asymmetric} setup). 
As illustrated in Table~\ref{tab:sr_beta_sym}, our approach generally works better in the symmetric setup (like other baseline methods). Still, Table~\ref{tab:sr_beta_asym} shows that our \textit{DRL-Agent} also dominates over competitors in the asymmetric settings. 
This quality makes our approach effective and applicable even when the underlying community detection algorithm is unknown. 


\subsection{Parameter Sensitivity}
\label{subsec:param-sensitivity}
The effectiveness of our \textit{DRL-Agent} relies on two critical parameters: $(i)$ the similarity threshold ($\tau$) used to determine whether the node-hiding goal has been achieved or not, and $(ii)$ the budget ($\beta$) to limit the effort -- i.e., graph modifications -- performed to achieve the goal. In this section, we analyze their impact.  
Specifically, in Table~\ref{tab:sr_kar_gre_tau-0.3}, we explore how the Success Rate for the community membership hiding task is influenced by varying the values of $\tau$ and $\beta$, while keeping the detection algorithm $f(\cdot)$ and dataset fixed. 
The results shown refer to a specific symmetric setup using the \texttt{greedy} community detection algorithm on the \texttt{words} dataset.
\begin{table}[htbp!]
    \caption{\label{tab:sr_kar_gre_tau-0.3}The impact of $\boldsymbol{\tau}$ and $\boldsymbol{\beta}$ on Success Rate (SR), using the \texttt{greedy} community detection algorithm on the \texttt{words} dataset.}
    \centering
    \scalebox{0.62}{
    \begin{tabular}{|c|c|c|c|c|c|c|c|}
        \hline
        \multirow{4}{*}{$\bm{\tau}$} & \multirow{4}{*}{$\bm{\beta}$} & \multicolumn{6}{c|}{\textbf{Community Membership Hiding Algorithm}} \\ 
        \cline{3-8} 
        & & \thead{DRL-Agent\\(ours)} & \textit{Random} & \textit{Degree} & \textit{Betweenness} & \textit{Roam} & \textit{Greedy} \\
        \hline
        \multirow{3}{*}{$0.3$} & $\frac{1}{2}\mu$ & $\mathbf{41.7\% \pm 5.6\%}$ & $39.0\% \pm 5.5\%$ & $39.3\% \pm 5.5\%$ & $40.7\% \pm 5.7\%$ & $36.7\% \pm 5.4\%$ & $26.7\% \pm 5.0\%$\\
         & $1\mu$ & $\mathbf{61.7\% \pm 5.5\%}$ & $46.0\% \pm 5.6\%$ & $53.7\% \pm 5.6\%$ & $55.0\% \pm 5.6\%$ & $41.3\% \pm 5.6\%$ & $54.7\% \pm 5.6\%$ \\
         & $2\mu$ & $\mathbf{77.7\% \pm 4.7\%}$ & $62.7\% \pm 5.5\%$ & $64.7\% \pm 5.4\%$ & $67.7\% \pm 5.3\%$ & $50.7\% \pm 5.7\%$ & $71.0\% \pm 5.1\%$ \\
        \hline
        \multirow{3}{*}{$0.5$} & $\frac{1}{2}\mu$ & $\mathbf{49.7\% \pm 5.7\%}$ & $48.7\% \pm 5.7\%$ & $48.0\% \pm 5.7\%$ & $46.0\% \pm 5.6\%$ & $45.3\% \pm 5.6\%$ & $29.0\% \pm 5.1\%$ \\
         & $1$$\mu$ & $\mathbf{71.7\% \pm 5.1\%}$ & $57.0\% \pm 5.6\%$ & $62.3\% \pm 5.5\%$ & $64.7\% \pm 5.4\%$ & $51.7\% \pm 5.7\%$ & $64.0\% \pm 5.4\%$ \\
         & $2\mu$ & $\mathbf{88.3\% \pm 3.6\%}$ & $71.0\% \pm 5.1\%$ & $77.3\% \pm 4.7\%$ & $79.7\% \pm 4.6\%$ & $44.7\% \pm 5.6\%$ & $79.7\% \pm 4.5\%$ \\
        \hline
        \multirow{3}{*}{$0.8$} & $\frac{1}{2}\mu$ & $\mathbf{62.7\% \pm 5.5\%}$ & $55.3\% \pm 5.6\%$ & $52.0\% \pm 5.7\%$ & $49.0\% \pm 5.7\%$ & $45.3\% \pm 5.6\%$ & $32.0\% \pm 5.3\%$\\
         & $1\mu$ & $\mathbf{90.0\% \pm 3.4\%}$ & $73.7\% \pm 5.0\%$ & $75.0\% \pm 4.9\%$ & $77.7\% \pm 4.7\%$ & $58.0\% \pm 5.6\%$ & $71.0\% \pm 5.1\%$ \\
         & $2\mu$ & $\mathbf{96.3\% \pm 2.1\%}$ & $82.3\% \pm 4.3\%$ & $86.3\% \pm 3.9\%$ & $88.3 \% \pm 3.6\%$ & $62.0\% \pm 5.5\%$ & $92.7\% \pm 2.9\%$\\
        \hline
    \end{tabular}
    }
\end{table}

As one might expect, by increasing the similarity threshold ($\tau$), the node deception goal is easier to achieve for our \textit{DRL-Agent} and, therefore, the Success Rate is higher (for a given fixed budget $\beta$). This happens because we pose a less strict requirement on the distance between the original community of the target node to hide and the one where it eventually ends up after the graph modifications induced by our method. 
Similarly, granting a larger budget ($\beta$) allows our method to alter more substantially the neighborhood of the target node to hide, hence increasing its chance of success. 

\subsection{Computational Complexity}
\label{subsec:complexity}
The size of the graph datasets used in our experiments ranges from small to large (e.g., the Facebook dataset \texttt{fb-75} has thousands of nodes and hundreds of thousands of edges). 
However, we acknowledge that real-world network graphs may count millions, if not billions, of nodes. 
Therefore, we analyze the computational complexity of our proposed method to assess its feasibility and potential deployment into extremely large-scale production environments.

The primary computational challenge of our method lies in training our \textit{DRL-Agent}. 
In each time step of the training process, the agent can select from a set of actions $\mathcal{A}$. Each action corresponds to either removing an existing edge or adding a new one, as detailed in Section~\ref{subsec:cf-graph}.
Thus, $|\mathcal{A}| \leq |\mathcal{V}|^2 = n^2$, considering the worst-case scenario of the target node being part of a fully connected graph.
The number of possible states ($\mathcal{S}$), instead, equates to the potential adjacency matrices for a graph with $|\mathcal{V}|=n$ nodes, i.e., $|\mathcal{S}| = 2^{n^2}$.

In general, given a discounted MDP $\mathcal{M} = \{\mathcal{S}, \mathcal{A}, \mathcal{P}, p_0, r, \gamma\}$ and assuming that sampling state-action pairs from the transition function $\mathcal{P}$ takes $O(1)$ time, \citet{neurips2018sidford} show that the upper bound on the time spent and number of samples taken for computing an $\epsilon$-optimal policy with probability $1-\delta$ is:
\[
O\Bigg[\frac{|\mathcal{S}||\mathcal{A}|}{(1-\gamma)^3\epsilon^2} \log\Bigg(\frac{|\mathcal{S}||\mathcal{A}|}{(1-\gamma)\delta \epsilon}\Bigg) \log\Bigg(\frac{1}{(1-\gamma)\epsilon} \Bigg)\Bigg].
\]
Given the exponential number of states in our worst case setting, our \textit{DRL-Agent} may result impractical to train for graphs with a large number $n$ of nodes.
Nonetheless, in practice, the number of states and actions allowed are significantly smaller. 
Experimental results, even on graphs with thousands of nodes like \texttt{fb-75}, demonstrate that convergence to the optimal policy occurs at a faster rate, and the theoretical complexity bound above might not be tight.

\section{Discussion}
\label{sec:discussion}

\subsection{Who Can Run Our Method?}
\label{subsec:implementation}
In our scenario, we assume that a social network runs the community detection algorithm $f(\cdot)$.
Moreover, this social network may offer its end-users the capability to opt out of being detected, accommodating their privacy needs. 
In such a case, the platform has \textit{full knowledge} of the graph, and our community membership hiding algorithm can be run seamlessly. 
Thus, the resulting counterfactual graph can be used to suggest to the target end-user what links they should add/remove to meet their deception goal.

On the contrary, if the social network does not offer this opt-out feature, the end-user may still attempt to remain concealed from the community detection algorithm. However, in this case, two critical considerations come into play, as the average end-user typically: $(i)$ lacks access to the full graph structure, and $(ii)$ cannot directly execute the community detection algorithm $f(\cdot)$.

To address $(i)$, the target end-user who wants to get masked off using our community membership hiding algorithm must first perform web scraping to obtain their local graph structure, albeit only the community to which they belong. 
For example, this community can be constructed by examining the followers of the people the target end-user follows. However, complications may arise, particularly on platforms like Instagram or Facebook, where viewing a user's followers is generally possible if you follow them back.

To tackle $(ii)$, the target end-user must establish empirical indicators that serve as proxies to determine if their hiding goal has been met.
For instance, this could involve criteria such as "not receiving any more following suggestions or advertisements that are clearly targeted for the community they want to be masked from."

Moreover, in a real-world scenario, multiple nodes may require concealment independently. There is a risk that modifications aiding the concealment of one node might hinder another's, potentially reducing the overall success rate. This issue arises when the nodes to be hidden share non-trivial neighborhood overlaps.
To address this, we propose extending our framework to handle node sets rather than individuals, akin to a multi-agent reinforcement learning (MARL) problem. Each agent, representing a node, collaborates to achieve their hiding objectives.

\subsection{Security and Ethical Implications}
\label{subsec:ethics}
As highlighted in the motivation for this work, community membership hiding algorithms can serve as valuable tools for safeguarding the privacy of social network users. 
Furthermore, these methods can be used to protect individuals at risk, including journalists or opposition activists, in regions governed by authoritarian regimes.
Additionally, these techniques can combat online criminal activities by modifying network connections to infiltrate espionage agents or disrupt communications among malicious users.

However, node-hiding techniques can also be exploited to pursue harmful goals.
For instance, malicious individuals can strategically use these methods to evade network analysis tools, often employed by law enforcement for public safety, enabling them to mask their illicit or criminal activities on the network.
Furthermore, in this work, we focus on a \textit{single} node $u$ to be hidden from a community without considering potential side effects on other nodes $v$. As we only manipulate edges controlled by $u$, the chance of another node $v$ unintentionally joining the community without any direct action from $v$ itself (e.g., adding a link to a node $w$ in the community) seems rare, albeit possible. To deal with such potential side effects, we can either limit the set of feasible actions or conduct post-processing sanity checks to ensure that no node that was not originally part of the community ends up included therein.

Overall, for a social network offering community membership hiding capabilities, it is essential to thoroughly assess the impact of this feature \textit{before} granting users the actual ability to conceal themselves from community detection algorithms.

\section{Conclusion and Future Work}
\label{sec:conclusion}
This paper tackled the challenge of \textit{community membership hiding}, which entails strategically modifying the structural characteristics of a network graph to prevent a target node from being detected by a community detection algorithm. 
To address this problem, we formulated it as a constrained counterfactual graph objective and solved it via deep reinforcement learning. 
We conducted extensive experiments to validate our method's effectiveness.
Results demonstrated that our approach strikes the best balance between achieving the desired node-hiding goal and the required cost of graph modifications compared to existing baselines.

In future work, we aim to explore different definitions of community membership hiding and incorporate node feature modifications alongside structural alterations to the counterfactual graph objective. 
Furthermore, we plan to apply our method to extremely large network graphs. 
Finally, we will generalize our method to address the community deception task or scenarios where multiple users simultaneously request node membership hiding. 

\begin{acks}
This work was partially supported by the projects FAIR (PE0000013), SERICS (PE00000014), and SoBigData.it (IR0000013) under the National Recovery and Resilience Plan funded by the European Union NextGenerationEU, as well as HypeKG -- Hybrid Prediction and Explanation with Knowledge Graphs (H53D23003700006) under the PRIN 2022 program funded by the Italian MUR.
\end{acks}

\bibliographystyle{ACM-Reference-Format}
\balance


\end{document}